\documentclass[11pt]{article}

\usepackage{graphics}
\input{fullpage.sty}

\begin{document}

\title{Accurate Calculations of the Peierls Stress in Small Periodic Cells}
\author{D.E. Segall \\ Department of Physics, Massachusetts Institute
of Technology, Cambridge MA 02139 \\
T.A. Arias \\ Laboratory of Atomic and Solid State Physics,
Cornell University, Ithaca, NY 14853  \\
Alejandro Strachan and William A. Goddard III\\ Materials and
Process Simulation Center, Beckman Institute (139-74) \\ California Institute of Technology, Pasadena, Ca 91125}

\date{\today}
\maketitle

\begin{abstract}	
The Peierls stress for a [111]-screw dislocation in bcc Tantalum is
calculated using an embedded atom potential. More importantly, a method
is presented which allows  accurate calculations of the Peierls
stress in the smallest periodic cells.
This method can be easily applied to {\em ab initio} calculations,
where only the smallest unit cells capable of containing a dislocation
can be conviently used. The calculation specifically focuses on the
case where the maximum resolved shear stress is along a $\{ 110
\}$-plane.
\end{abstract}


\section{Introduction}

Long, low-mobility [111]-screw dislocations control
low temperature plastic behavior in bcc metals. Unlike their fcc counter
parts, bcc metals violate Schmid behavior and have many active slip
planes. The microscopic origins of such behavior is crucial in the
understanding of their plastic behavior. Therefore detailed and
accurate first principle calculations are invaluable.

The Peierls stress, the zero temperature limit of the critical stress
for slip, has been calculated in a variety of ways using various types
of boundary
conditions~\cite{wang,shenoy,moriarty,xu,takeuchi,yang,bulatov,rao1}.
These calculations, in general, rely on empirical potentials to
simulate various configurations and incorporate a relatively large
number of atoms. In general three types of boundary conditions are
used, fixed cylindrical boundary
conditions~\cite{shenoy,moriarty,xu,takeuchi}, lattice greens
functions~\cite{rao2,yang,rao1} or periodic boundary
conditions~\cite{wang,bulatov}. For these approaches a large number of
atoms is generally taken in order to minimize any artificial effects
the boundary conditions may impose.  Generally, this number is too
large to be suited for {\em ab initio} calculations, even on the
fastest and largest super-computers.  Applying corrections to these
effects is therefore important in order to have a reliable {\em ab
initio} calculation.
	
The simplest approach is to use fixed cylindrical boundary conditions.
However, care must be taken due to the mismatch of the boundary with
the dislocation, particularly when the dislocation moves.  This
mismatch can be minimized by performing calculations in very large
systems, effectively extrapolating to the limit of infinite cylinder
size.  The size of the system needed to extract accurate values can be
greatly reduced by applying leading order corrections due to the
mismatch in the boundary~\cite{shenoy}. However, even with these
corrections, accurate calculations can generally be made only for a
cylinder with a radius of $\sim $30$\AA$ or greater, which corresponds
to $\sim $700 atoms when using periodic boundary conditions along the
dislocation line.  Such fixed cylindrical boundary conditions are
generally ill-suited for an {\em ab initio} calculation due to both
this large number of atoms and the artificial effects of the free
surface at the boundary on the electrons.  An alternative approach to
account for boundary effects, and thereby reduce the size of the
system needed for accurate calculation, is to use lattice greens
function techniques~\cite{yang,rao1,rao2}.  This technique, although
quite elegant, still faces the issue of free surface effects when
applied to first principles quantum mechanical calculations.

Periodic boundary conditions are in general the most natural and
straight forward way to calculate the Peierls stress in a density
functional theory calculation. In periodic boundary conditions one is
forced to have a net zero Burgers vector per unit cell. This is
generally accomplished by using either a dipole or quadrupole array of
dislocations~\cite{marklund,bigger,lehto}.  Peierls stress
calculations have been done in periodic boundary conditions using
empirical potentials, however, to avoid dislocation-dislocation
interactions, generally requires the use of unit cells~\cite{bulatov}
which would be impracticable for {\em ab initio} calculations.  Wang
{\em et al.}\cite{wang} have introduced a promising method for use in
small, periodic cells, but this approach requires the decomposition of
the total energy into individual atomic contributions, which while
feasible for inter-atomic potential approaches, is not well-defined in
quantum mechanical calculations.

In this paper it is shown how an accurate Peierls stress can be
obtained in very small periodic cells. It is shown how leading order
effects, due to the array of closely packed dislocations, can be
accurately accounted.  It is further shown how an accurate Peierls
stress can be obtained by simply applying a pure shear to the system
and minimizing the energy.  This is a tremendous advantage in {\em ab
initio} calculations as it minimizes the number of atomic
configurations which need be explored.  As an application of these
ideas, we focus specifically on the calculation of the Peierls stress
in bcc tantalum for a [111]-screw dislocation in which the maximum
resolved shear stress is along a $\{ 110 \}$.  This is among
the most relevant geometries in understanding the plastic response of
this material.

The paper is organized as follows.  Section~2 presents the results of
Peierls stress calculations in very large cylindrical cells, as a
reference against which we shall compare our new approach.  Section~3
presents the results of calculations using periodic boundary
conditions and underscores the problems that arise.
Section~4 shows how to overcome these problems, and, finally,
Section~5 presents several techniques for accelerating the calculation.
Finally, Section~6 concludes the paper.

\section{Reference Calculations} \label{sec:cylinder}

All calculations presented throughout this manuscript employ a
quantum-based, embedded atom force field (qEAMFF) developed
in~\cite{strachan}.  To provide the reference value of the Peierls
stress for our calculations within periodic boundary conditions, we
first perform calculations for isolated dislocations in a cylindrical
geometry with fixed boundary conditions.

\begin{figure}
\begin{center}
\scalebox{0.85}{\includegraphics{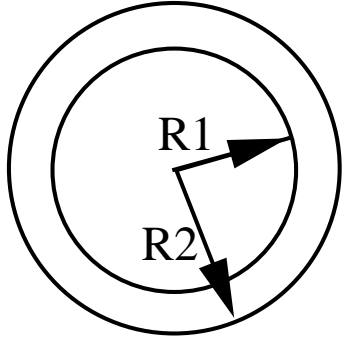}}

\end{center}                                        
\caption{Peierls stress calculation within cylindrical boundary
conditions: atoms whose distance from the center is less than $R1$ are
relaxed under inter-atomic potential forces, while those in the region
between $R1$ and $R2$ are held fixed to the anisotropic elasticity
theory solution. }
\label{fig:cyl}                                    
\end{figure}

To form the dislocation, we proceed as follows.  First, we define
coordinates for our calculations as follows: the $x$-axis lies along
the $[1\bar{1}0]$ direction, the $y$ lies along the $[11\bar{2}]$, and
the $z$ lies along the $[111]$. A cylinder with radius $R2$ of Ta is
then taken (Figure~\ref{fig:cyl}) which has two regions, a fixed
region and a relaxation region. In the relaxation region ($r < R1$),
where $r$ is the radial distance from the center, the atoms are
allowed to relax in response to their interatomic forces. The atoms in
the fixed region ($R1 < r < R2$) are held fixed to their positions
according to the solution of anisotropic elasticity theory.  Finally,
periodic boundary conditions (with period of one Burger's vector) are
employed along the direction of the dislocation.  Thus, these
calculations consider only straight, infinite dislocations. A
[111]-screw dislocation then is obtained by displacing all atoms in
the cylinder according to the solution of anisotropic elasticity
theory~\cite{stroh,head} and then optimizing the positions of the
atoms in the relaxation region.

\begin{figure}
\begin{center}
\scalebox{0.5}{\includegraphics{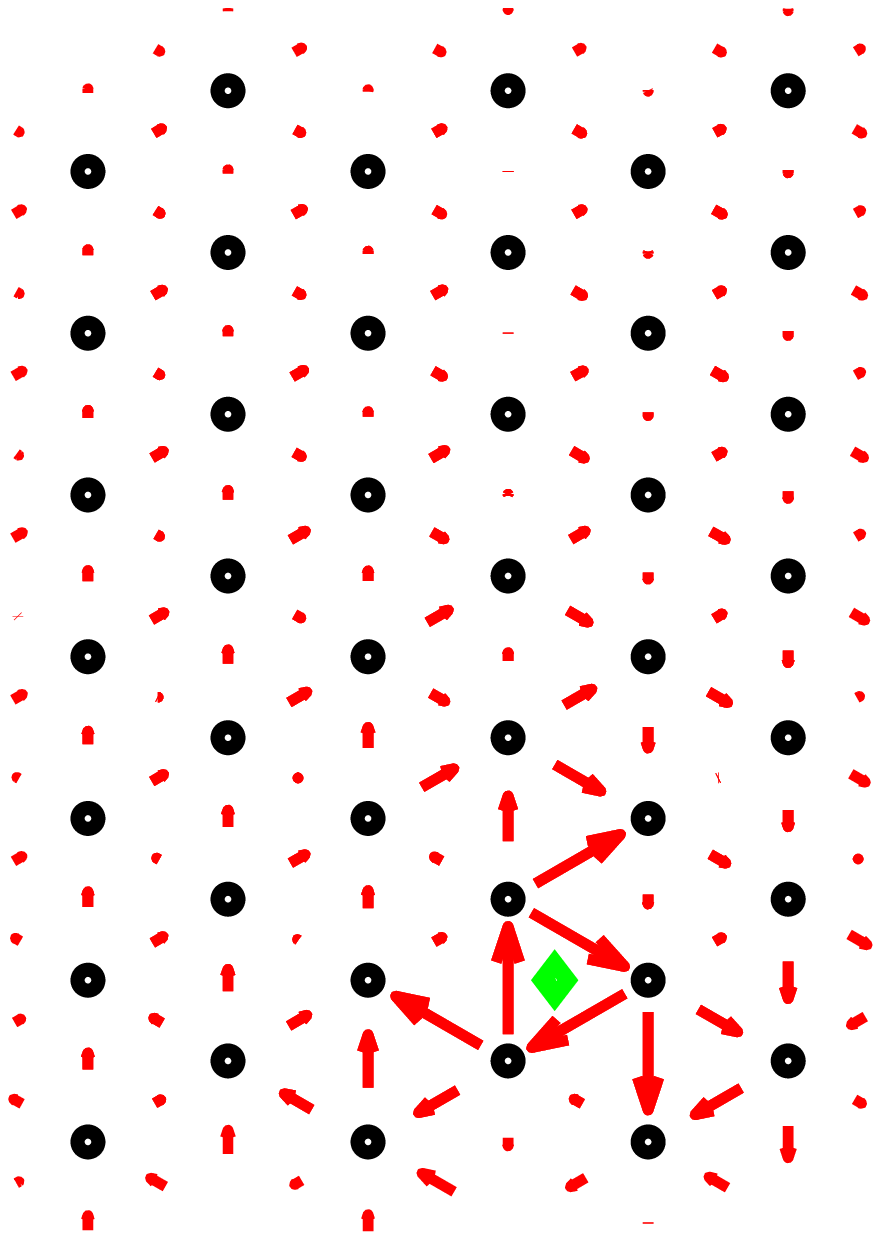}}  \hspace{1.15cm}
\scalebox{0.5}{\includegraphics{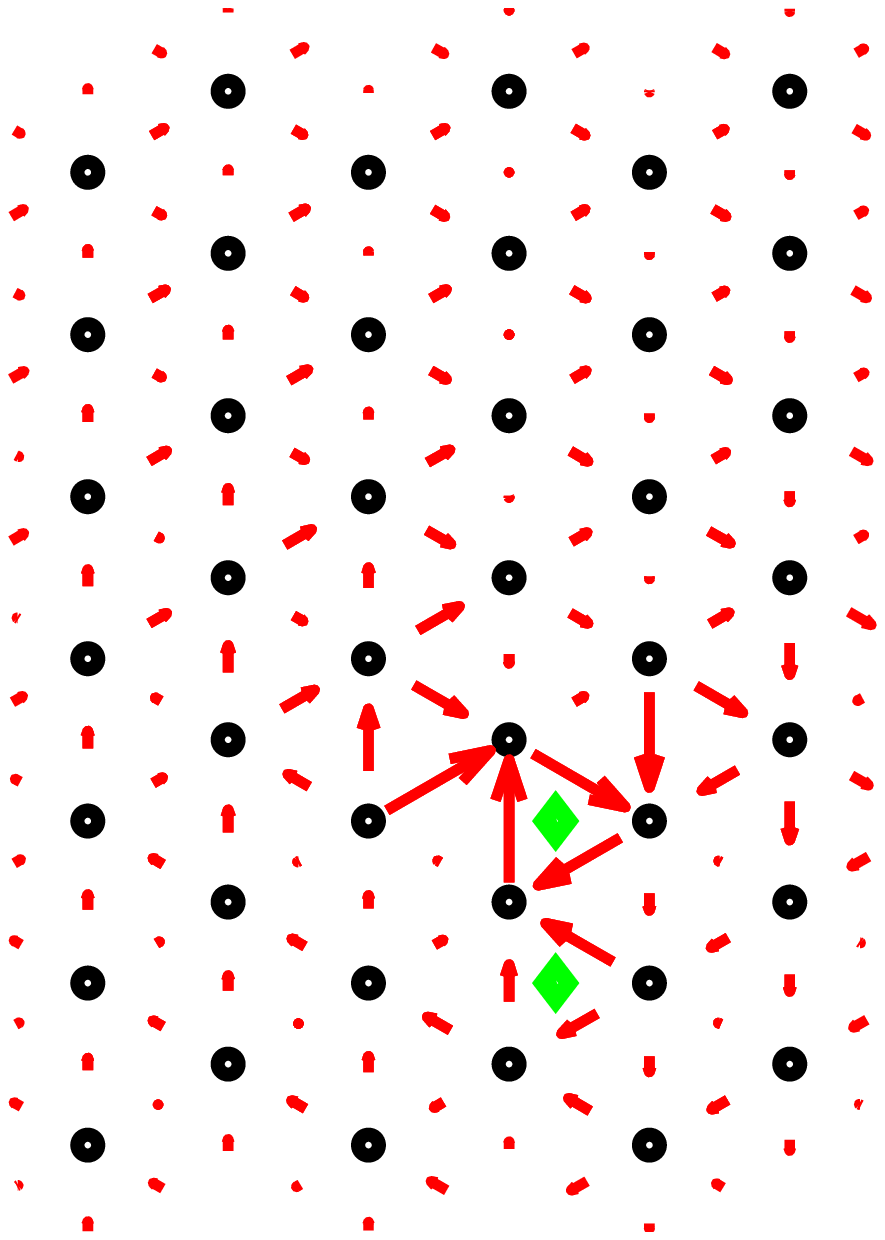}}  \hspace{1.15cm}
\scalebox{0.5}{  \includegraphics{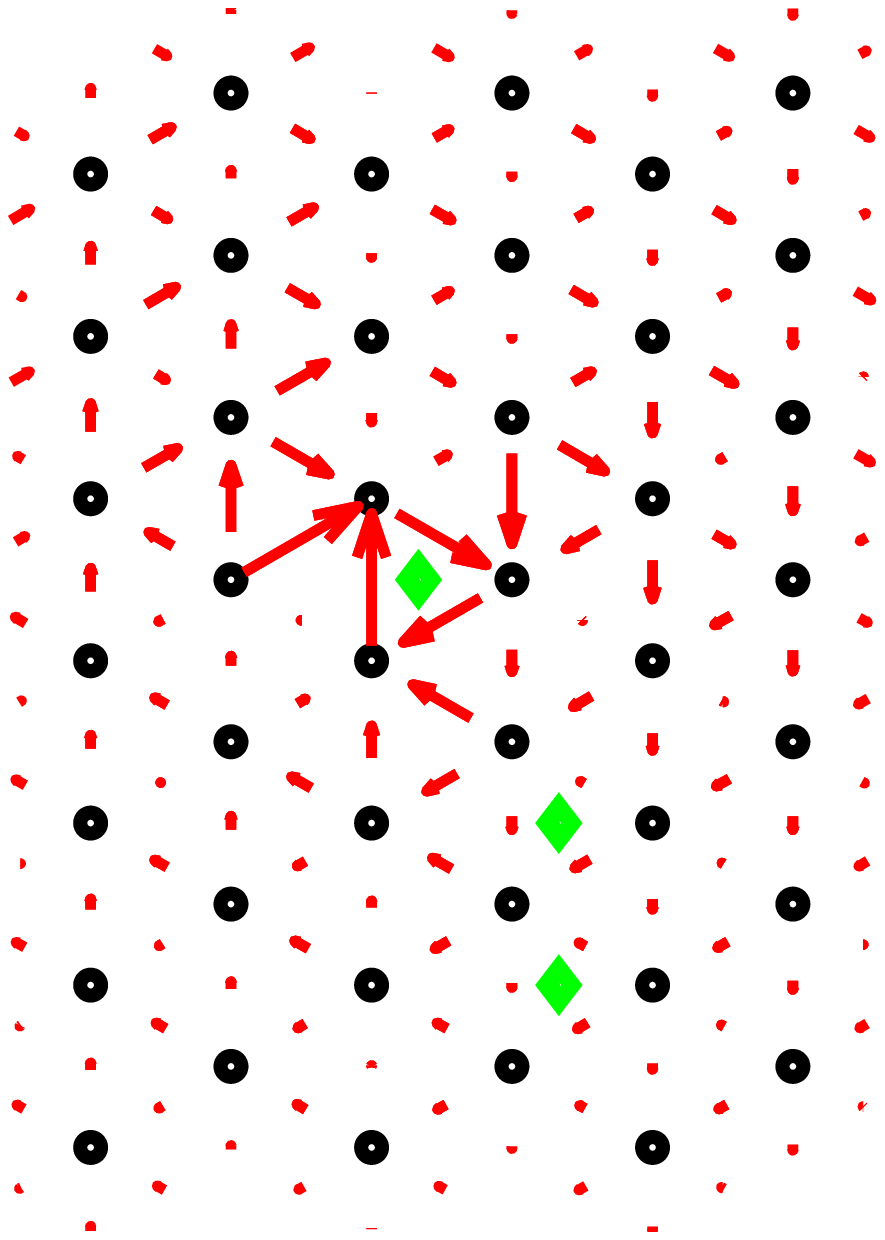}}
\\  \ \\
  {\LARGE{\bf (a)} \hspace{1.9in} {\bf (b)} \hspace{1.9in} {\bf (c)}} 
\\ \ \\ \ \\
 \scalebox{0.5}{\includegraphics{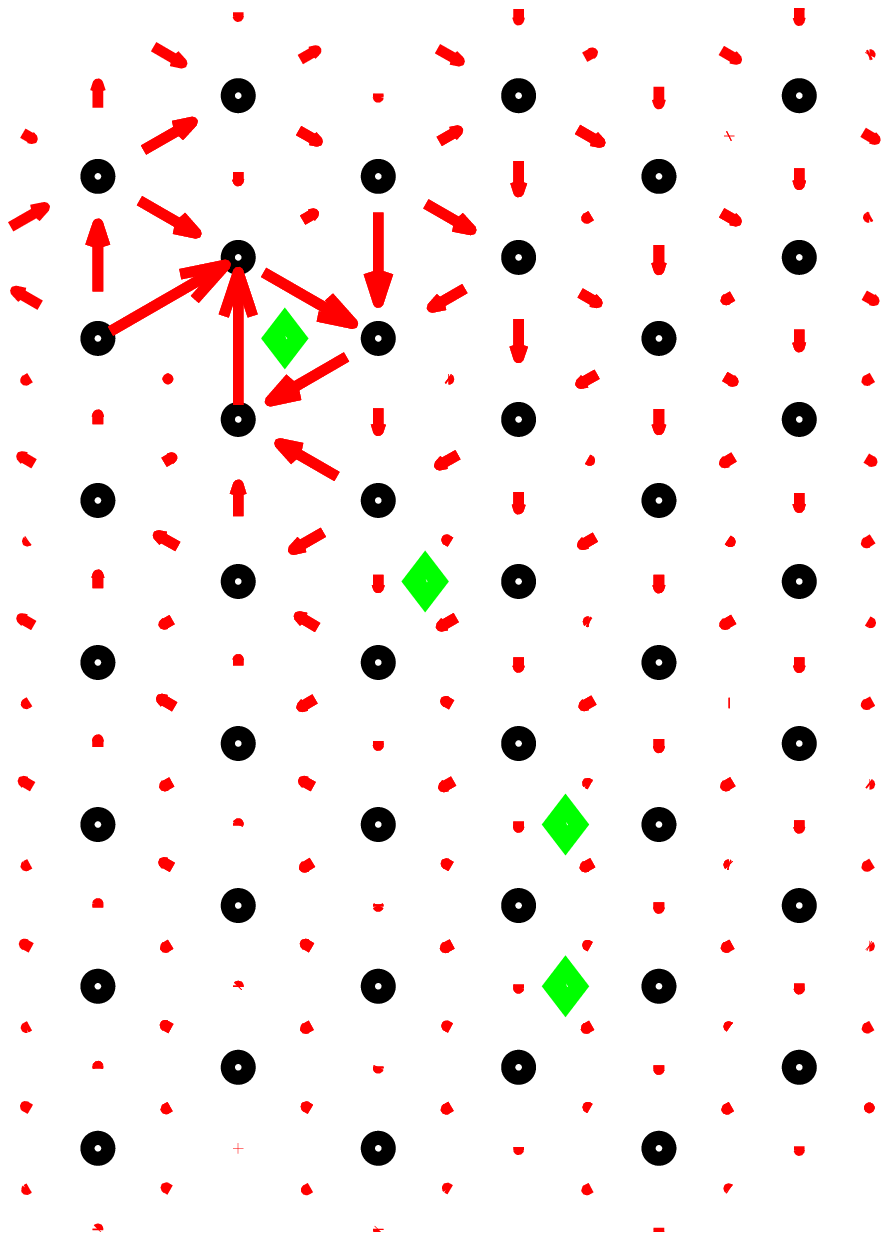}}  \hspace{4cm}
\scalebox{0.5}{\includegraphics{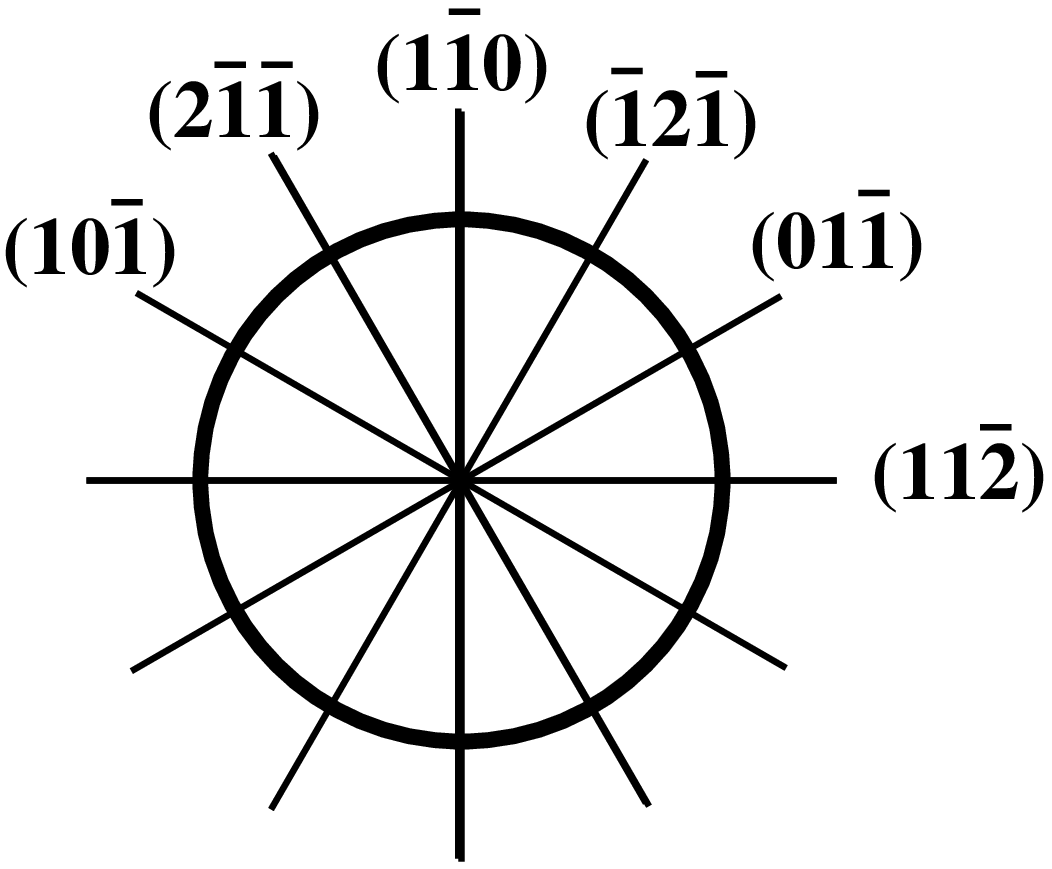}}
\\ \ \\
 {\LARGE {\bf (d)} \hspace{2.8in} {\bf (e)}  }
\end{center}                                        
\caption{Differential displacement (DD) maps of a [111] screw
dislocation at increasing stress computed within cylindrical boundary
conditions: {\bf a}) zero applied stress, {\bf b}) initial jump along
$(1\bar{1}0)$ at a stress $\geq $ 0.74 Gpa ($P1$), {\bf c}) second
jump, along $(2\bar{1}\bar{1})$ at a stress $\geq $ 0.91 GPa
($P2$), {\bf d)} subsequent jump along $(2\bar{1}\bar{1})$, {\bf e)}
orientation of \{110\} and \{112\} planes. }
\label{fig:disloc}                                    
\end{figure}

To calculate the Peierls stress, we then apply a strain to the system
which ensures that the resulting stress has only one component,
$\sigma_{xz}$, which generates a force on the dislocation line in the
$[11\bar{2}]$ direction~\cite{hirth}.  We then optimize the locations
of the atoms in the relaxation region subject to this external strain.
Figure~\ref{fig:disloc} shows the resulting dislocation structure in
a large cyllinder ($R1 \approx 120\AA$) for a series of different
strains using a differential displacement (DD) map~\cite{vitek}.  In
these maps, the circles represent columns of atoms viewed along the
$[111]$ direction.  The arrows indicates the change in relative
displacement that neighboring atomic columns make, relative to the
bulk, due to the presence of the dislocation.  The lengths of the
arrows are normalized so that a displacement of 1/3 of a Burgers
vector corresponds to an arrow of full length.  The first panel (a)
shows the ground-state structure of the dislocation at zero stress.
The center of the dislocation is located at a diamond encased in triad
of arrows.  Going around this triad makes three displacements of
one-third of a Burgers vector for a net displacement of one full
Burgers vector relative to the bulk.  The ground state structure of
the core is seen to break the symmetry of the lattice by extending
outward along three $\{110\}$ planes.  This ``degenerate split core''
is consistent with molecular dynamic results, using periodic boundary
conditions, found in reference~\cite{wang} when using the same
interatomic potential.

As the strain is applied to the ground state structure
(Figures~\ref{fig:disloc}b-d), the dislocation feels force along the
$(1\bar{1}0)$-plane.  As the applied strain obtains a critical value,
the dislocation center moves one lattice spacing along the
$(1\bar{1}0)$-plane (Figure~\ref{fig:disloc}b).  As the strain is
increased further, the dislocation then glides along the
$(2\bar{1}\bar{1})$ (Figures~\ref{fig:disloc}~c,d) in two steps, first
along $(10\bar{1})$ and then spontaneously along $(1\bar{1}0)$.  There
are consequentially two Peierls barriers that the dislocation must
overcome, the first along $(1\bar{1}0)$ and the second along
$(10\bar{1})$.  This leaves some ambiguity for the definition of the
Peierls stress.  Different authors have used different
definitions~\cite{wang,xu,takeuchi,yang}.  For clarity we will consider
the value of the $\sigma_{xz}$ stress for the first jump
(Figure~\ref{fig:disloc}a-b) as the first Peierls stress ($P1$) and
the value of the $\sigma_{xz}$ stress for the second jump
(Figure~\ref{fig:disloc}b-c) as the second Peierls stress ($P2$).

\begin{figure}
\begin{center}
\rotatebox{90}{{\LARGE \hspace*{.9in} {\bf Critical Shear} ${\mathbf (\sigma_{xz})}$\bf{ [GPa]}}}
\scalebox{0.85}{\includegraphics{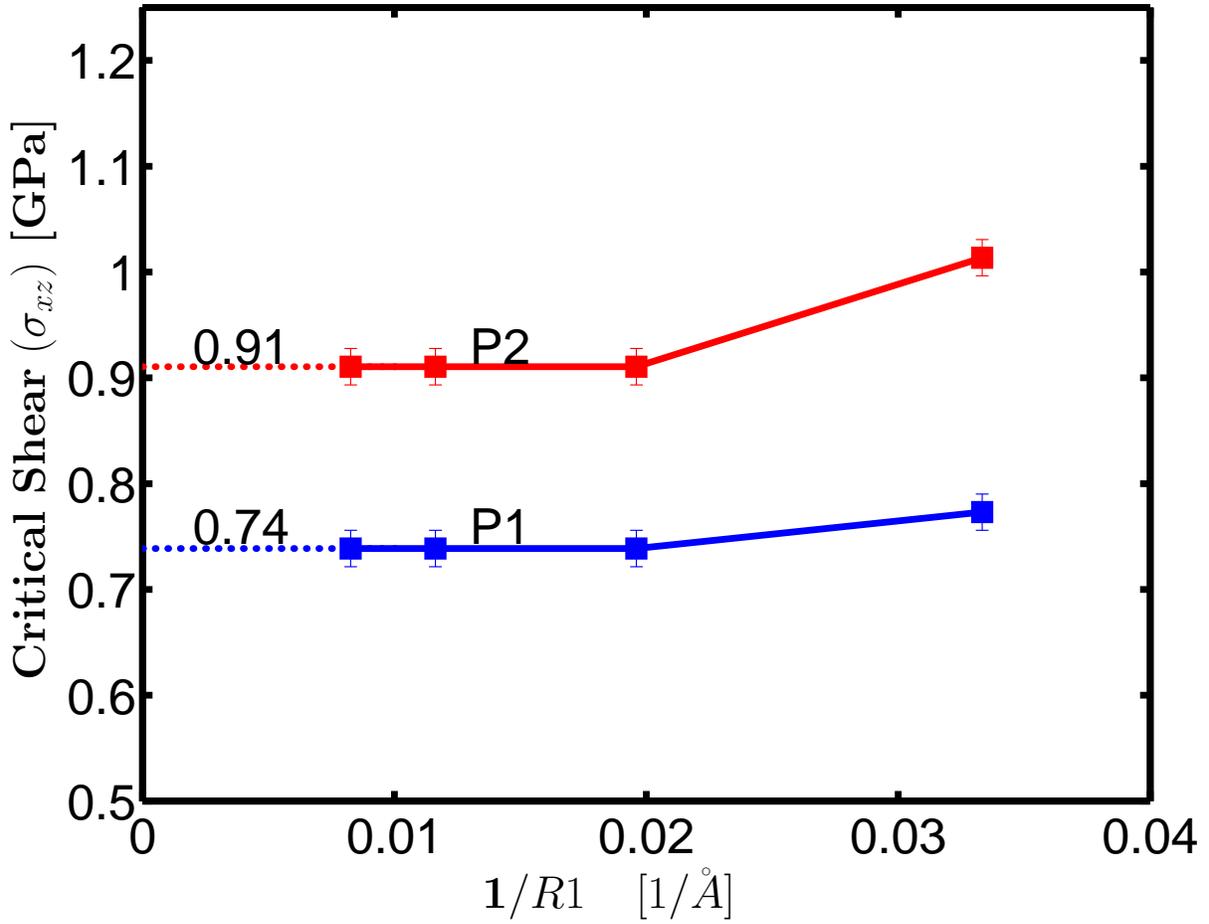}} \\
{\LARGE ${\mathbf 1/R1}$ \hspace{.3cm} ${\mathbf [1/\AA]}$}
\end{center}                                        
\caption{Convergence of P1 and P2 as a function of $1/R1$, for cylinders
ranging from $R1=30\AA$ to $R1=120\AA$. }
\label{fig:Cylinf}                                    
\end{figure}

To extract the limit of these critical stresses for an isolated
dislocation in an infinite crystal, we have repeated the above
calculations in cylinders of various sizes ranging from $R1=30\AA$ to
$R1=120\AA$ and extracted the Peierls stress as a function of radius.
Figure~\ref{fig:Cylinf} summarizes our results.  In the limit, we find
values of 0.74 GPa and 0.91 GPa for $P1$ and $P2$, respectively.

As a consistency check that we indeed expect finite-size effects to be
small for our largest cylinders, one can employ the
method of Shenoy and Phillips~\cite{shenoy}. This methods estimates
the unaccounted restoring stress that the boundary applies to a
displaced dislocation to be 
$$
\frac{K_{s} b}{2\pi}A\frac{d}{R^2},
$$
where $K_{s} = (S_{11}/(S_{11}S_{44} - S_{15}^{2}))^{1/2}$ in terms of
the modified elastic compliances $S_{ij}$~\cite{stroh}, $b$ is the
Burgers vector, $d$ is the distance the dislocation has moved from the
center of the cylinder, $R$ is the radius of the cylinder, and $A$ is
a dimensionless constant that can be calculated through elasticity
theory or computationally.  In our case, $A\approx 2$, $K_{s}=68$GPa
and $b = 2.9\AA$.  Finally, for $P1$ we have $d\approx .1\AA$ and for
$P2$ we have $d\approx 2.7\AA$.  For our largest cylinder ($R1 =
120\AA$), we calculate a restoring stress of $0.0004$GPa and
$0.012$GPa, respectively.  Both values are quite small, well within
the uncertainties in Figure~\ref{fig:Cylinf}.

\section{Periodic Boundary Conditions}

\begin{figure}
\begin{center}
\rotatebox{90}{{\LARGE \hspace*{1.75in} {\bf $[1 1 \bar{2}]$}}}
\scalebox{0.85}{\includegraphics{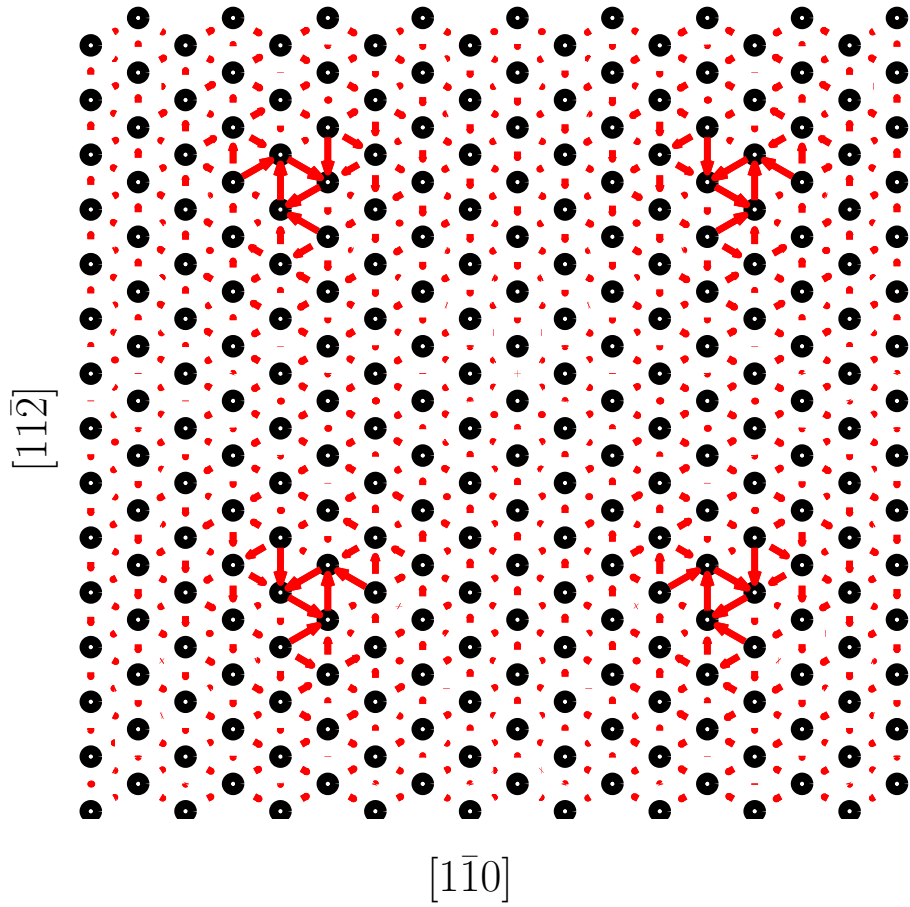}} \\ \ \\
{\LARGE {\bf $[1 \bar{1} 0]$}}
\end{center}                                        
\caption{Dislocation displacement (DD) map of a quadrupole dislocation
array within a 270 atom cell with periodic boundary conditions.  The
cell may be reduced to 135 atoms with appropriate choice of lattice
vectors.}
\label{fig:quad}                                    
\end{figure}

In periodic boundary conditions, the unit cell must contain a net zero
Burgers vector.  In practice this is generally accomplished through
the use of either a dipolar or quadrupolar array of
dislocations~\cite{marklund,bigger,lehto}.  Our calculations employ a
quadrupole array, which has been shown to be the more appropriate
choice for screw-dislocations~\cite{lehto}.  Figure~\ref{fig:quad}
illustrates a cell containing 270 atoms.

As our first attempt to calculate the Peierls stress in periodic
boundary conditions, we take a quadrupole array with lattice vectors
fixed at the values corresponding to the perfect bulk material, that
is the appropriate lattice vectors prior the insertion of
dislocations.  Below, we refer to this choice as ``unrelaxed'' lattice
vectors, as they generally do not correspond to the lattice vectors of
an unstrained quadrupole array.

To extract an estimate of the Peierls stress, we then follow a
procedure analogous to that in Section~\ref{sec:cylinder}, applying
strain until the dislocations move.  Here, rather than applying the
strain to the fixed region ($R1<r<R2$), we strain the {\em unrelaxed}
lattice vectors.  Then, we then compute the critical stress from the
strain through the elastic constant matrix,
\begin{eqnarray}
{\mathbf \sigma} & = & {\mathbf C}\cdot {\mathbf \epsilon}, \label{eqn:sigeps}
\end{eqnarray}
where ${\mathbf \sigma}$ is a column vector of the stresses,
${\mathbf \epsilon}$ is the applied strain vector, and ${\mathbf C}$ is
the elastic constant matrix, {\em assumed} to equal that of the bulk
material.  The advantage of this approach is that it requires
exploration of a minimal number of configurations and bulk lattice and
elastic constants are relatively easy to obtain from first principles
calculations.  The disadvantage of this approach is that its underlying
assumptions cast doubt of the accuracy of the results for unit cells
of modest size.

\begin{figure}
\begin{center}
\rotatebox{90}{{\LARGE \hspace*{.5in} {\bf Critical Shear} ${\mathbf (\sigma_{xz})}$\bf{ [GPa]}}}
\scalebox{0.75}{\includegraphics{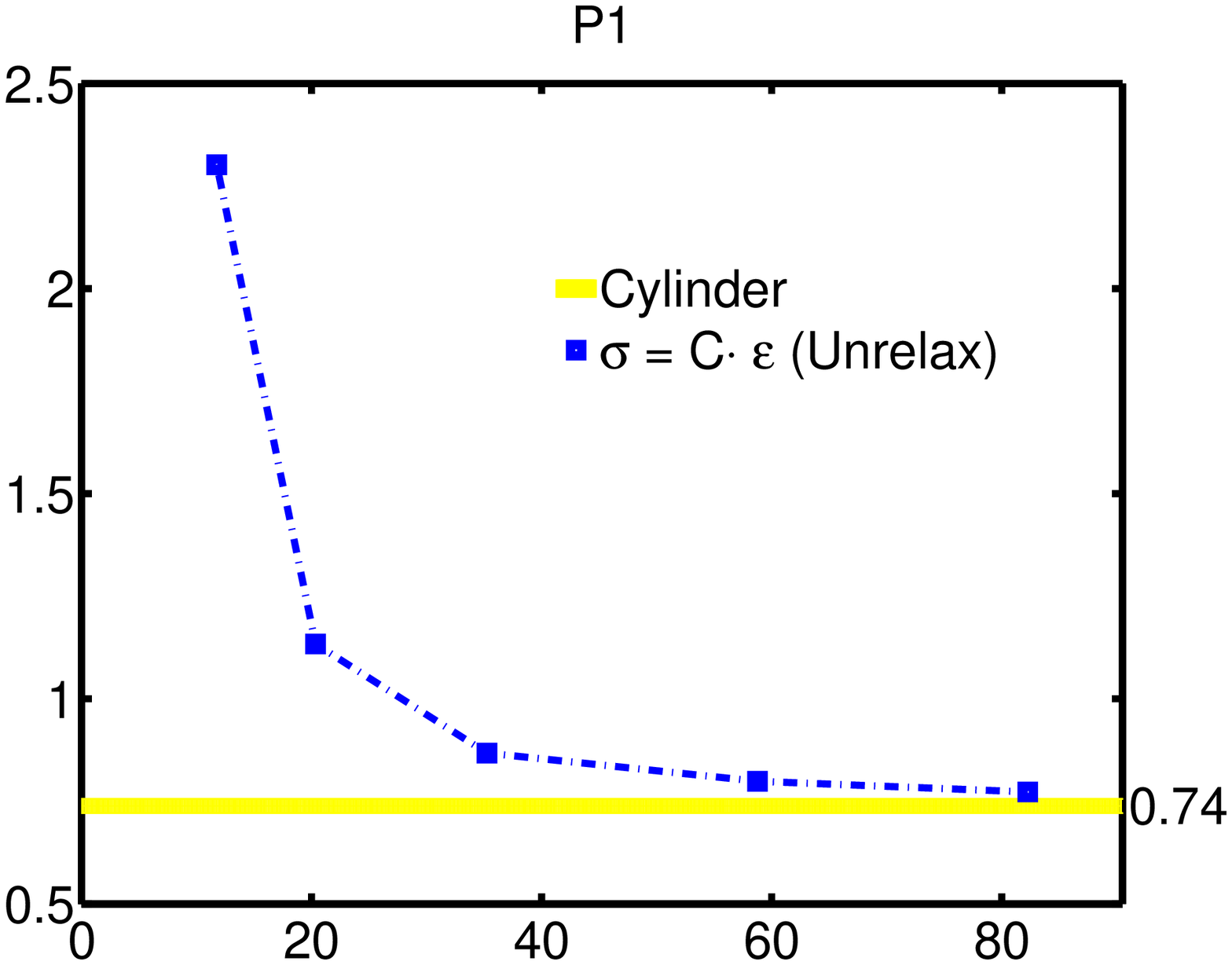}}\\
{\LARGE \bf{Dislocation Separation} ${\mathbf [\AA]}$}

\end{center}                                        
\caption{Convergence of P1 calculated within periodic boundary
conditions using unrelaxed lattice vectors and bulk elastic constants:
Peierls stress calculated using Eq.~\ref{eqn:sigeps} (squares), limit
exacted from Figure~\ref{fig:Cylinf} (solid line with width indicating
numerical uncertainty).  Uncertainties in the periodic calculations
are smaller than the square symbols.}
\label{fig:bulkel}                                    
\end{figure}

Figure~\ref{fig:bulkel} explores the convergence of this approach with
increasing cell size for calculation of the first Peierls stress
($P1$).  The two smallest cells contain 90 atoms ($\sim 2.9\AA \times
24\AA \times 24 \AA$) and 270 atoms ($\sim 2.9\AA \times 42\AA \times
42 \AA$) (or 45 and 135 atoms, if symmetry is exploited), and are the
only cells convenient for detailed {\em ab initio} studies.  The
results in these cells, however, are extremely poor, with errors of
$200\%$ and $53\%$, respectively.  Thus, great care must be taken when
working with such small cells and a method is needed to correct for
these finite-size effects.

\section{Corrections}

The preceding calculations make two major assumptions: (1) that the
quadrupole array of dislocations does not change the elastic constant
matrix from that of bulk material, and (2) that the use of unrelaxed
lattice vectors is appropriate.  To explore the impact of these
assumptions, we repeat the above calculations while using both the
relaxed lattice vectors and elastic constants of the quadrupolar
array.

Numerical calculations show that the symmetry of the elastic constant
matrix of the quadrupolar array (${\mathbf C'}$) is the same as that of
the bulk, although the individual components may differ.  In
particular, the stress-strain relation takes the form
\begin{eqnarray}
{\mathbf \sigma} & = & {\mathbf C'}\cdot {\mathbf \epsilon} \label{Cmat} 
\end{eqnarray}
\begin{eqnarray}
\left[ \begin{array}{c} \sigma_{xx} \\ \sigma_{yy} \\
\sigma_{zz} \\ \sigma_{yz} \\ \sigma_{xz} \\ \sigma_{xy}
 \end{array} \right]& = & {\Huge  \left[ \begin{array}{cc}  \left(4\times
4\right) & 0 \\   0   & \left(2\times 2\right) \end{array} \right]}
\cdot  \left[ \begin{array}{c}  \epsilon_{xx} \\ \epsilon_{yy} \\
\epsilon_{zz} \\ \epsilon_{yz} \\ \epsilon_{xz} \\ \epsilon_{xy}
 \end{array} \right]    \label{eqn:str-str}
\end{eqnarray}
In this equation, the lower $2 \times 2$ sub-block couples the
$\sigma_{xz}$ and $\sigma_{xy}$ stresses only to the corresponding
strains, while the upper $4 \times 4$ block couples the $\sigma_{xx},
\sigma_{yy}, \sigma_{zz}$ and $\sigma_{yz}$ stresses only to their
corresponding strains.  Note, therefore, that for the present Peierls
stress calculations only the lower $2 \times 2$ block is relevant.

\begin{figure}
\begin{center}
\rotatebox{90}{{\LARGE \hspace*{.5in} {\bf Critical Shear} ${\mathbf (\sigma_{xz})}$\bf{ [GPa]}}}
\scalebox{0.75}{\includegraphics{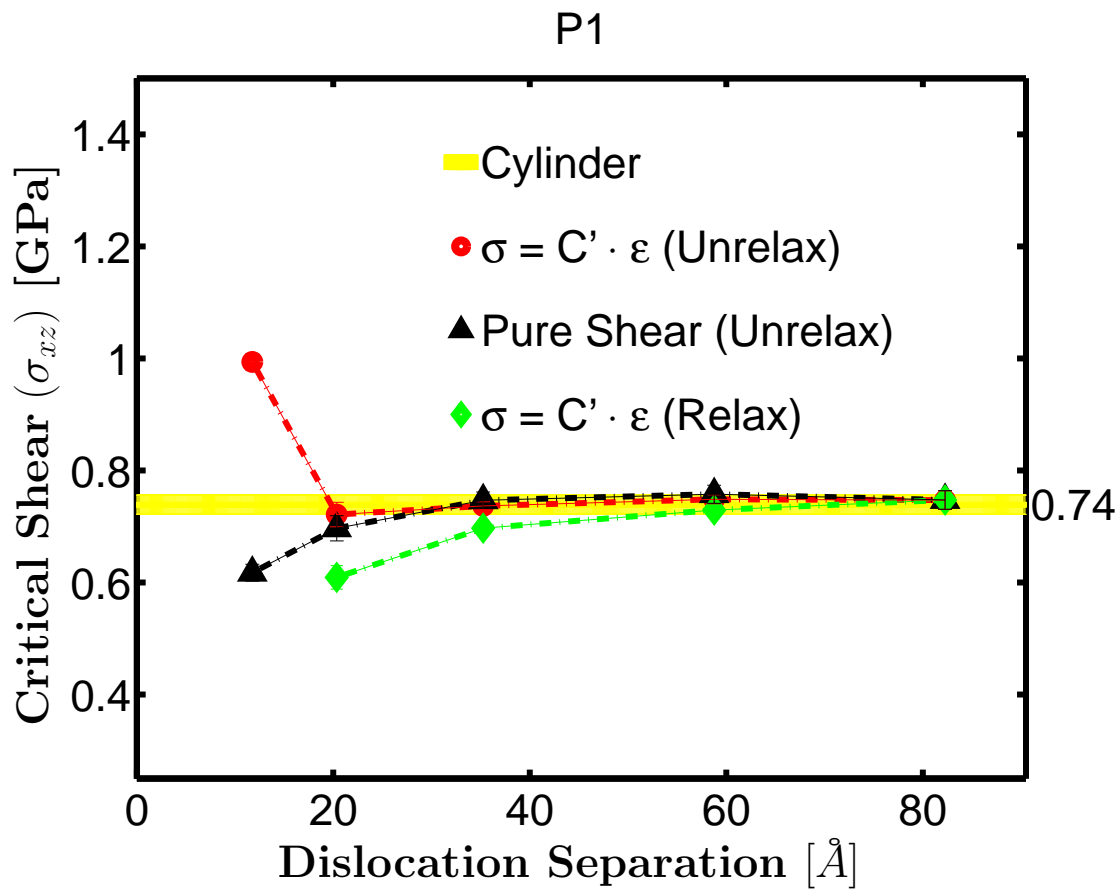}}\\
{\LARGE \bf{Dislocation Separation} ${\mathbf [\AA]}$}
\end{center}                                        
\caption{Convergence of P1 calculated within periodic boundary
conditions: using relaxed lattice vectors and quadrupolar elastic
constants (diamonds), using unrelaxed lattice vectors and quadrupolar
elastic constants (circles), using unrelaxed lattice vectors, pure
shear and elastic constants extracted during the calculation
(triangles), asymptotic result extracted from cylindrical boundary
conditions (horizontal line). Error bars associated with quadrupolar
array are generally smaller than the associated icons.}
\label{fig:pureshear}                                    
\end{figure}

Figure~\ref{fig:pureshear} shows the results (diamonds in the
figure) of the extraction of the Peierls stress when using relaxed
lattice vectors and the elastic constants for the quadrupole.  The
figure shows that the Peierls stress now converges much more quickly
with cell size.  (The figure does not include results for the smallest
cell, which proved unstable to the relaxation of the lattice vectors.)
For the smallest stable cell (135 atoms including symmetry), the error
is reduced from $53\%$ to only $18\%$, indicating the possibility of
extracting reasonable results from cells of size suited to {\em ab
initio} calculations.  Although results for reasonably sized cells are
accurate, this approach, however, is not necessarily well suited for
{\em ab initio} calculations because relaxation of the lattice vectors
and the calculation of the elastic constant matrix requires the
exploration of many new atomic configurations.

\section{Accurate and Efficient Peierls Stress Calculation}

Having found an accurate approach, we now explore how to minimize that
calculations associated with the above corrections.  We begin by
considering the necessity of relaxing the lattice vectors and then
consider computation of the {\em relevant} components of the elastic
constant matrix ${\mathbf C'}$.

\subsection{Benefit of fixed lattice vectors} ~\label{sec:norellats}

Working with unrelaxed lattice vectors generates spurious strains in
the unit cell.  The question is whether these strains create spurious
stresses which confound the extraction of the Peierls stress.
The elastic force which any such stresses would generate on the
dislocation take the form~\cite{hirth}:
\begin{equation}
{\mathbf F}_{L} = {\mathbf b}\cdot {\mathbf \sigma}\times {\mathbf \eta}.
\label{eqn:pkf}
\end{equation}
Here, ${\mathbf F}_{L}$ is the force per unit length on the
dislocation, ${\mathbf b}$ is the Burgers vector, ${\mathbf \sigma} $ is
the stress tensor written as a {\em matrix}, and ${\mathbf \eta}$ is
the sense vector, the direction along which the dislocation runs.  For
a $[111]$-screw dislocation ${\mathbf b}$ and ${\mathbf \eta}$ both lie
along the $z-$axis, and, therefore, only two components of stress can
generate a force on the dislocation, $\sigma_{xz}$ and $\sigma_{yz}$,
and confound the extraction of the Peierls stress.  We now consider
whether the strains associated with the unrelaxed lattice vectors can
generate such stresses.  We distinguish two types of strain,
dilation and shear.

The presence of the quadrupolar array tends to dilate the unit cell
in the (111) plane.  Due to symmetry, this dilation tends to be
uniform, as we have confirmed by direct numerical calculation on the
unit cells.  The above form of the elastic constant matrix ${\mathbf C'}$
prevents such dilation from generating any $\sigma_{xz}$ component.
The second confounding component $\sigma_{yz}$ would also vanish were
the $4 \times 4$ subblock of the elastic constant matrix to have
precisely the same symmetry as that of the bulk.  Numerically, we find
that that this is almost the case, and that the dilation contribution
to $\sigma_{yz}$ ranges from $0.02\%$ (in our largest cell) to only
$4\%$ (in the 135 atom cell) of the stress experienced in the {\em
unrelaxed} cell.  The diagonal components of strain
$\epsilon_{ii}$ do not significantly affect the extraction of the
Peierls stress, and therefore need not be relaxed to extract
meaningful Peierls stresses.

\begin{figure}
\begin{center}
\scalebox{0.75}{\includegraphics{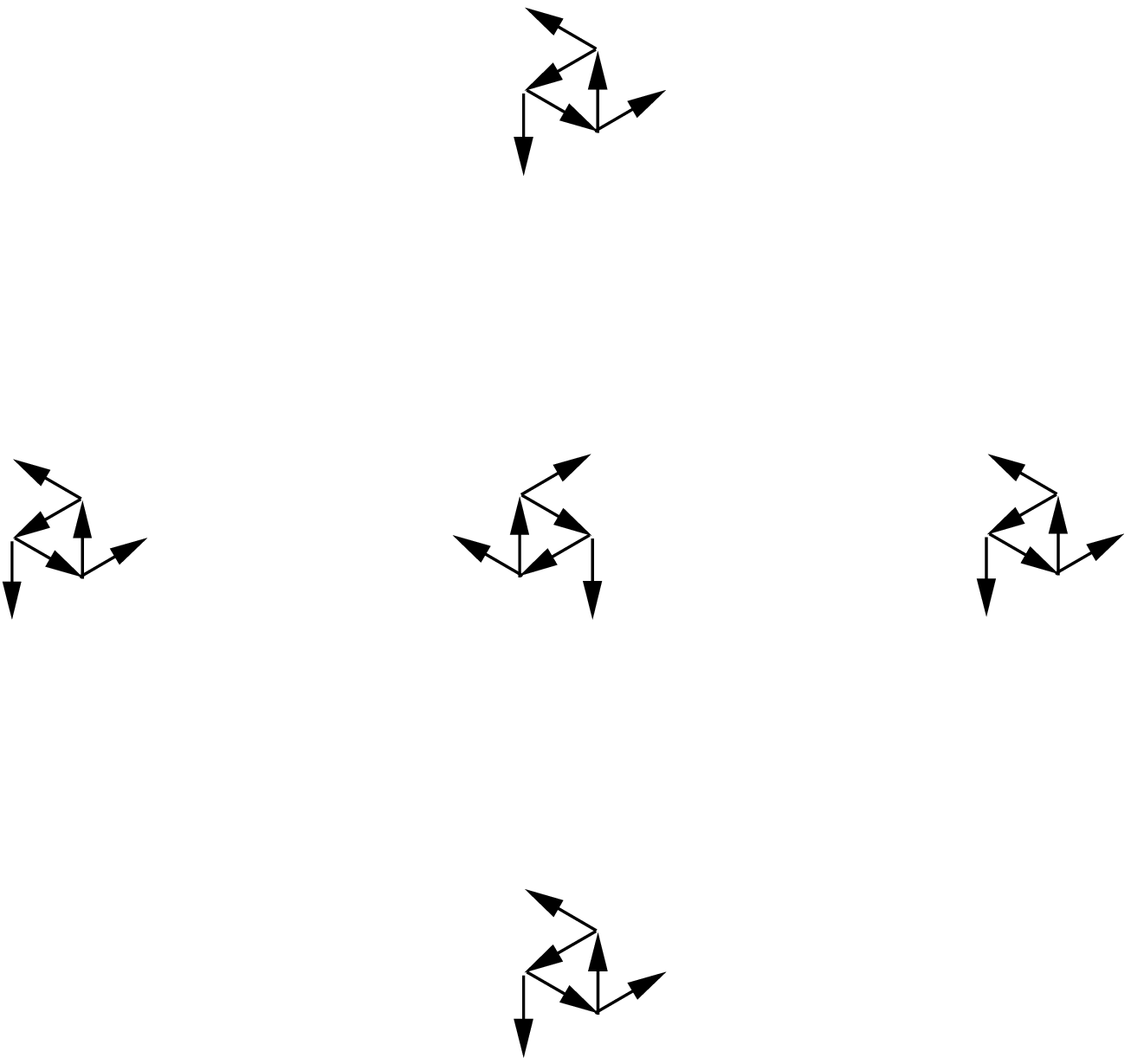}}
\end{center}
\rotatebox{90}{\LARGE $\longrightarrow [11\bar{2}]$}
{\LARGE  $\longrightarrow [1\bar{1}0]$}                                        
\caption{The first nearest neighbors cannot generate any elastic force
on the center dislocation, since they are equally spaced. From the
core asymmetry there can be a core-core force in only in the
$[1\bar{1}0]$. Note that these symmetry arguments works for higher
order neighbors too.}
\label{fig:coreasym}                                    
\end{figure}

Turning to shear strains, the equal spacing of the dislocation array
leads to zero shear within linear elasticity theory.  From symmetry,
however, core-core interaction (non-elastic) can generate a force in
the $[1\bar{1}0]$-direction only, Figure~\ref{fig:coreasym}, which
results in one non-zero component of strain, $\epsilon_{yz}$. From
direct calculations it is shown that this is in fact the only non-zero
component of shear strain.  Relaxing this component of strain,
therefore, creates a artificial material environment, different from
what isolated dislocations would experience.  To more quickly approach
the limit of isolated dislocations, one should therefore {\em not}
relax this component of the strain.  This advantage of using of
unrelaxed lattice vectors is well known and commonly exploited in the
materials literature in the study of {\em two}-dimensional defects
such as grain boundaries or surfaces\cite{arias,jwang}.  We therefore
expect to be able to extract accurate Peierls stresses without the
need for relaxing either off-diagonal or diagonal components of
strain.

Figure~\ref{fig:pureshear} shows the results (circles in the
figure) of extracting the Peierls stress when using {\em unrelaxed}
lattice vectors, but while still computing the stresses with the
appropriate elastic constant matrix ${\mathbf C'}$.  The results
converge even more quickly than those obtained by relaxing the lattice
vectors, thus supporting our analysis.  Therefore, extremely good
results (already within $2\%$ in the 135 atom cell) can be obtained by
not relaxing the lattice vectors.  This not only produces results of
far superior quality, it also reduces the computational effort.

\subsection{Extraction of elastic constants}

The results of Section~\ref{sec:norellats}, while quite impressive,
still require calculation of the elastic constant matrix ${\mathbf
C'}$ of the quadrupolar array.  Significant savings can be gained with
the simple realization that only the $2\times2$ sub-block of the
elastic constant matrix enters calculation of $P1$ and $P2$ through
Eq.~\ref{eqn:str-str}.  Moreover, within linear elasticity theory,
these components can be extracted equally well at the relaxed or
unrelaxed lattice vectors, thereby again mitigating the need for
relaxation of the lattice vectors.

Not all of the components of the $2 \times 2$ sub-block,
\begin{equation}
{\mathbf C'}_{2} \equiv \left[ \begin{array}{cc} C'_{44} & -C'_{16} \\
 					   -C'_{16} & C'_{55}
			\end{array} \right], \label{eqn:C2}
\end{equation}
need be computed independently.

\begin{figure}
\begin{center}
\rotatebox{90}{\hspace*{1.2in} \LARGE{${\mathbf \Delta}$\bf{ E [eV]}}}
\scalebox{0.75}{\includegraphics{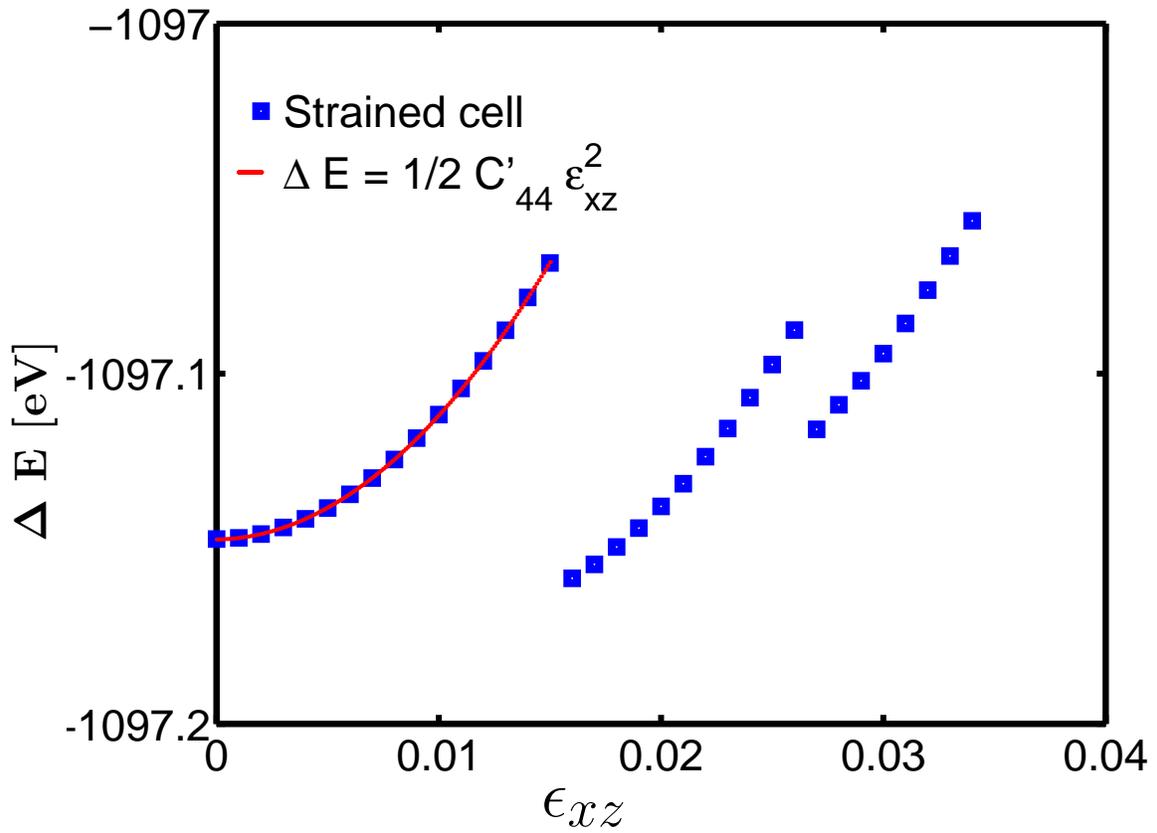}}\\
{\Huge ${\mathbf \epsilon_{xz}}$}
\end{center}                                        
\caption{Energy of a $42\AA \times 42\AA \times 2.9\AA$ cell (270
atoms) plotted as a function of applied pure shear strain
$\epsilon_{xz}$: direct calculations (squares), quadratic fit
(curve).}
\label{fig:energy}                                    
\end{figure}

To see this, consider the application of a pure, shear strain
$\epsilon_{xz}$ to the system.  As a result of the symmetry of ${\mathbf C'}$,
this generates only two components of stress, $\sigma_{xz}$ and
$\sigma_{xy}$.  From equation~\ref{eqn:pkf}, however, $\sigma_{xy}$ does
not generate any forces on the dislocation and thus, within linear
elasticity, does not effect our calculation of the Peierls stress.
The remaining stress, whose critical value is the Peierls stress, can be
computed from just one component of ${\mathbf C'}_2$,
\begin{equation}
\sigma_{xz} = C'_{44}\epsilon_{xz}. \label{eqn:newcoeff}
\end{equation}
Finally, we note that this pre-factor can be extracted without any
additional calculation.  Figure~\ref{fig:energy} shows the total
energy of the quadrupole array plotted as a function of strain during
the extraction of the critical Peierls stress.  Prior to the first
dislocation glide event at $\epsilon_{xz} \approx 0.015$, the energy
increases quadratically according to
\begin{equation}
\Delta E = \frac{1}{2}C'_{44}\epsilon_{xz}^{2}, \label{eqn:dE}
\end{equation}
which contains precisely the same pre-factor as in Eq.~\ref{eqn:newcoeff}.

Figure~\ref{fig:pureshear} shows the result (triangles in the
figure) of the extraction of the Peierls stress from the application
of a pure, shear strain $\epsilon_{xz}$ to the {\em unrelaxed} lattice
vecotrs and extracting the relevant elastic constant from the
 curvature of the energy prior to the glide events.  The results are of
nearly the same quality as working with unrelaxed lattice vectors and
using the full elastic constant matrix ${\mathbf C'}$.

\subsection{Second Peierls stress: $P2$}

\begin{figure}
\begin{center}
\rotatebox{90}{{\LARGE \hspace*{.5in} {\bf Critical Shear} ${\mathbf (\sigma_{xz})}$\bf{ [GPa]}}}
\scalebox{0.75}{\includegraphics{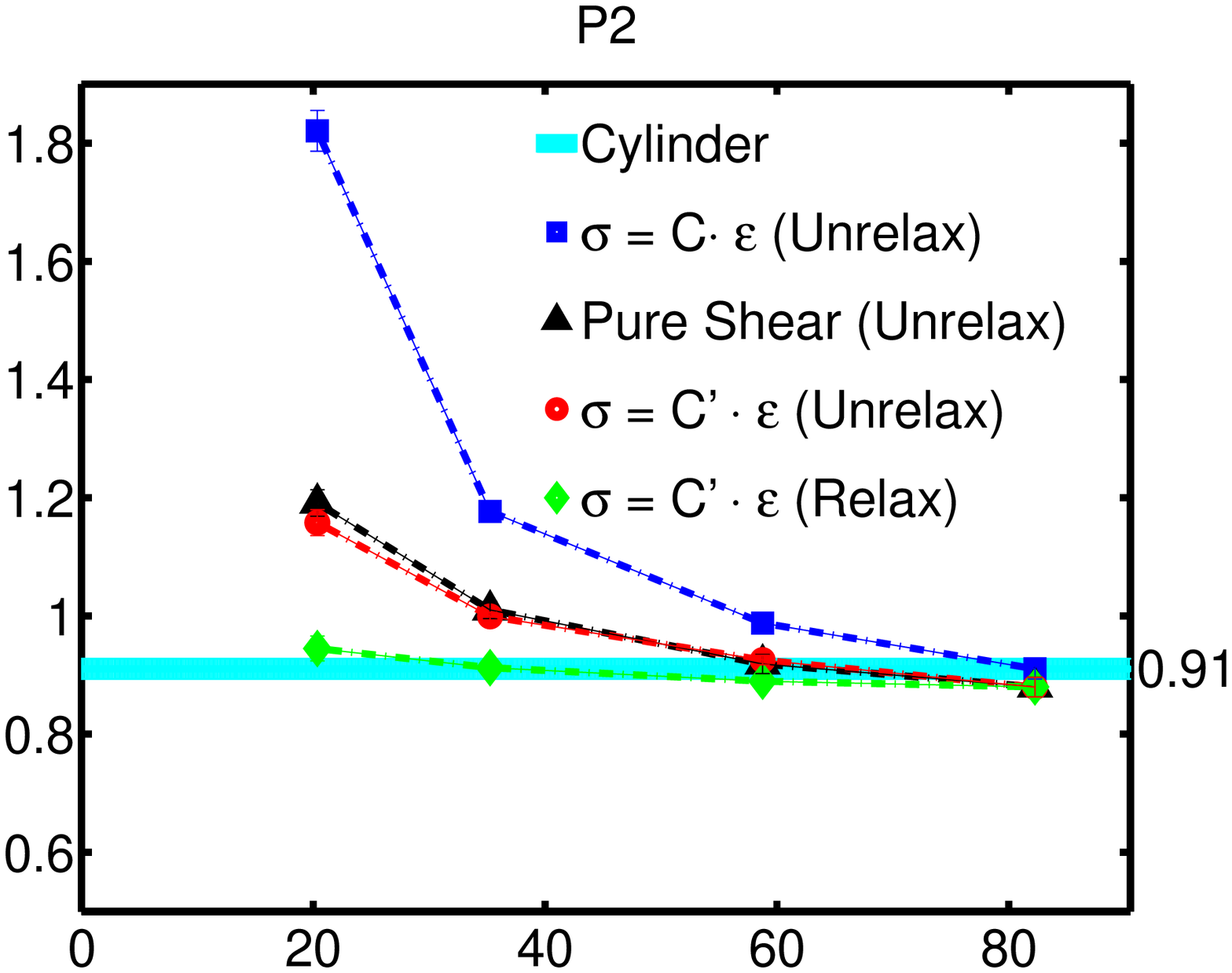}}\\
{\LARGE \bf{Dislocation Separation} ${\mathbf [\AA]}$}
\end{center}                                        
\caption{Convergence of P2 calculated within periodic boundary
conditions: using unrelaxed lattice vectors and bulk elastic constants
(squares), using relaxed lattice vectors and quadrupolar elastic
constants (diamonds), using unrelaxed lattice vectors and quadrupolar
elastic constants (circles), using unrelaxed lattice vectors, pure
shear and elastic constants extracted during the calculation
(triangles), asymptotic result extracted from cylindrical boundary
conditions (horizontal line).}
\label{fig:P2}                                    
\end{figure}

Figure~\ref{fig:P2} shows our preliminary results for the the
extraction of the second Peierls stress $P2$.  These results are
complicated by the fact that after the first transition $P1$, the
quadrupole array is now distorted and no longer a perfect quadrupole,
thereby generating non-negligible dislocation-dislocation forces and
further modifying the elastic constant matrix of the cell.  As a
possible correction to this effect, we are presently exploring
moving the distorted cores back to their original quadrupole locations
before further increasing the stress.  Nonetheless, we find that
significant improvements can be made by using the elastic constant
matrix of a perfect quadrupole array and ignoring the
dislocation-dislocation interactions.  Finally, results for pure shear
calculations are quite encouraging.

\section{Conclusion}

This paper present an accurate and effective way to calculate the
$\{110\}$ Peierls stress in a $[111]$ screw dislocation in a bcc
material.  The results show that accurate results can be obtained even
for the smallest cells while using unrelaxed lattice vectors and
extracting the elastic constants directly from the calculations.  The
method most importantly appears to make {\em ab initio} calculations
of Peierls stresses viable in periodic boundary conditions for the
first time.

\bibliographystyle{unsrt}
\bibliography{jcamd}

\begin{thebibliography}{10}

\bibitem{wang}
G.~Wang, A.~Strachan, T.~Cagin, and W.A. GoddardIII.
\newblock {\em Mater. Sci. and Engng.}, 2001.

\bibitem{shenoy}
V.~J. Shenoy and R.~Phillips.
\newblock {\em Phil. Mag. A}, 76:367, 1997.

\bibitem{moriarty}
J.~A. Moriarty, W.~Xu, P.~S\"{o}derlind, L.~H. Yang, and J.~Zhu.
\newblock {\em J. Engng. Mater. Technol.}, 121:120, 1999.

\bibitem{xu}
W.~Xu and J.~A. Moriarty.
\newblock {\em Comp. Mater. Sci.}, 9:348, 1998.

\bibitem{takeuchi}
S.~Takeuchi.
\newblock Core structure and glide behavior of a screw dislocation in teh bcc
  lattice.
\newblock In J.~K. Lee, editor, {\em Interatomic Potentials and Crystalline
  Defects}, page 201, 1980.

\bibitem{yang}
L.~H. Yang, S\"{o}derlind P., and J.~A. Moriarty.
\newblock {\em Phil. Mag. A} future issue.

\bibitem{bulatov}
V.~V. Bulatov, O.~Richmond, and M.~V. Glasov.
\newblock {\em Acta Mater.}, 47:3507, 1999.

\bibitem{rao1}
S.~Rao and C.~Woodward.
\newblock {\em Phil. Mag. A} future issue.

\bibitem{rao2}
S.~Rao, C.~Hernandez, J.~Simmons, T.~Parthasarathy, and C.~Woodward.
\newblock {\em Phil. Mag. A}, 77:231, 1998.

\bibitem{marklund}
Marklund S.
\newblock {\em Phys. Status Solidi B}, 85:673, 1978.

\bibitem{bigger}
J.~R.~K. Bigger and {\em et. al.}
\newblock {\em Phys. Rev. Lett.}, 69:2224, 1992.

\bibitem{lehto}
N.~Lehto and S.~Oberg.
\newblock {\em Phys. Rev. Lett.}, 80:5568, 1998.

\bibitem{strachan}
A.~Strachan, T.~Cagin, O.~Gulseren, S.~Mukherjee, R.~E. Cohen, and W.~A.
  Goddard~III.
\newblock {\em In preperation}.

\bibitem{stroh}
A.~N. Stroh.
\newblock {\em Phil. Mag.}, 3:625, 1958.

\bibitem{head}
A.~K. Head.
\newblock {\em Phys. Stat. Sol.}, 6:461, 1964.

\bibitem{hirth}
J.~P. Hirth and J.~Lothe.
\newblock {\em Theory of Dislocations}.
\newblock John Wiley and Sons, 2 edition, 1982.

\bibitem{vitek}
V.~Vitek.
\newblock {\em Cryst. Lattice Defects}, 5:1, 1974.

\bibitem{arias}
T.A. Arias and J.D. Joannopoulos.
\newblock {\em Phys. Rev. Lett.}, 69:3330, 1992.

\bibitem{jwang}
J.~Wang, T.A. Arias, and J.D. Joannopoulos.
\newblock {\em Phys. Rev. B}, 47:10497, 1993.

\end{thebibliography}

\newpage

\newpage

\newpage 

\newpage

\newpage

\newpage

\newpage

\newpage

\newpage

\end{document}